\documentclass[runningheads]{llncs}
\usepackage[T1]{fontenc}
\usepackage{booktabs}
\usepackage{amsmath}
\usepackage{amssymb}
\usepackage{graphicx}
\usepackage{multicol}
\usepackage{multirow}
\usepackage{adjustbox}
\usepackage{hyperref}
\hypersetup{
    colorlinks=true,
    linkcolor=blue,
    filecolor=magenta,      
    urlcolor=cyan,
    pdftitle={Single-subject Multi-contrast MRI Super-resolution via INRs},
    pdfpagemode=FullScreen,
    }
\usepackage{orcidlink}
\usepackage{url}

\begin{document}
\title{Single-subject Multi-contrast MRI Super-resolution via Implicit Neural Representations}
\titlerunning{Single-subject Multi-contrast MRI Super-resolution via INRs}

%

\author{Julian McGinnis\thanks{equal contribution}\inst{1,2,3}\orcidlink{0009-0000-2224-7600} \and
Suprosanna Shit$^\star$\inst{1,4,5}\orcidlink{0000-0003-4435-7207} \and
Hongwei Bran Li \inst{5}\orcidlink{0000-0002-5328-6407} \and Vasiliki~Sideri-Lampretsa\inst{1}\orcidlink{0000-0003-0135-7442} \and Robert Graf\inst{1,4}\orcidlink{0000-0001-6656-3680} \and Maik Dannecker\inst{1}\orcidlink{0000-0001-9012-9606} \and Jiazhen~Pan\inst{1}\orcidlink{0000-0002-6305-8117} \and Nil Stolt-Ansó\inst{1}\orcidlink{0009-0001-4457-0967} \and Mark Mühlau\inst{2,3}\orcidlink{0000-0002-9545-2709} \and Jan S. Kirschke\inst{4}\orcidlink{0000-0002-7557-0003} \and Daniel~Rueckert\inst{1,6}\orcidlink{0000-0002-5683-5889} \and Benedikt Wiestler\inst{4}\orcidlink{0000-0002-2963-7772}
} 

%
\authorrunning{McGinnis and Shit et al.}

\institute{
School of Computation, Information and Technology, TU Munich, Germany
\and
TUM-Neuroimaging Center, TU Munich, Germany 
\and
Department of Neurology, TU Munich, Germany 
\and
Department of Neuroradiology, TU Munich, Germany 
\and
Department of Quantitative Biomedicine, University of Zurich, Switzerland
\and
Department of Computing, Imperial College London, United Kingdom
\\
\email{\{julian.mcginnis,suprosanna.shit\}@tum.de}}

\maketitle              
\begin{abstract}

Clinical routine and retrospective cohorts commonly include multi-parametric Magnetic Resonance Imaging; however, they are mostly acquired in different anisotropic 2D views due to signal-to-noise-ratio and scan-time constraints.
Thus acquired views suffer from poor out-of-plane resolution and affect downstream volumetric image analysis that typically requires isotropic 3D scans.
Combining different views of multi-contrast scans into high-resolution isotropic 3D scans is challenging due to the lack of a large training cohort, which calls for a subject-specific framework.
This work proposes a novel solution to this problem leveraging Implicit Neural Representations (INR).
Our proposed INR jointly learns two different contrasts of complementary views in a continuous spatial function and benefits from exchanging anatomical information between them.
Trained within minutes on a single commodity GPU, our model provides realistic super-resolution across different pairs of contrasts in our experiments with three datasets.
Using Mutual Information (MI) as a metric, we find that our model converges to an optimum MI amongst sequences, achieving anatomically faithful reconstruction.\footnote{Code is available at: \url{https://github.com/jqmcginnis/multi_contrast_inr/}}

\end{abstract}

\section{Introduction}
In clinical practice, Magnetic Resonance Imaging (MRI) provides important information for diagnosing and monitoring patient conditions \cite{menze2014multimodal,commowick2016msseg}. To capture the complex pathophysiological aspects during disease progression, multi-parametric MRI (such as T1w, T2w, DIR, FLAIR) is routinely acquired.
Image acquisition inherently poses a trade-off between scan time, resolution, and signal-to-noise ratio (SNR) \cite{plenge2012super}.
To maximize the source of information within a reasonable time budget, clinical protocol often combines anisotropic 2D scans of different contrasts in complementary viewing directions.
Although acquired 2D scans offer an excellent in-plane resolution, they lack important details in the orthogonal out-of-plane.
For a reliable pathological assessment, radiologists often resort to a second scan of a different contrast in the orthogonal viewing direction.
Furthermore, poor out-of-plane resolution significantly affects the accuracy of volumetric downstream image analysis, such as radiomics and lesion volume estimation, which usually require isotropic 3D scans.
As multi-parametric isotropic 3D scans are not always feasible to acquire due to time-constraints~\cite{plenge2012super}, motion~\cite{gholipour2010robust}, and patient's condition~\cite{ha2020one}, super-resolution offers a convenient alternative to obtain the same from anisotropic 2D scans. Recently, it has been shown that acquiring three complementary 2D views of the \emph{same contrast} may yield higher SNR at reduced scan time \cite{plenge2012super,wu2021irem}. However, it remains under-explored if orthogonal anisotropic 2D views of \emph{different contrasts} can benefit from each other based on the underlying anatomical consistency. Additionally, whether such strategies can further decrease scan times while preserving similar resolution and SNR remains unanswered. Moreover, unlike conventional super-resolution models trained on a cohort, a personalized model is of clinical relevance to avoid the danger of potential misdiagnosis caused by cohort-learned biases. In this work, we mitigate these gaps by proposing a novel multi-contrast super-resolution framework that only requires the patient-specific low-resolution MR scans of different sequences (and views) as supervision. As shown in various settings, our approach is not limited to specific contrasts or views but provides a generic framework for super-resolution. The contributions in this paper are three-fold:

\begin{enumerate}
    \item To the best of our knowledge, our work is the first to enable subject-specific multi-contrast super-resolution from low-resolution scans without needing any high-resolution training data. We demonstrate that Implicit Neural Representations (INR) are good candidates to learn from complementary views of multi-parametric sequences and can efficiently fuse low-resolution images into anatomically faithful super-resolution.
    \item We introduce Mutual Information (MI) \cite{wells1996multi} as an evaluation metric and find that our method preserves the MI between high-resolution ground truths in its predictions. Further observation of its convergence to the ground truth value during training motivates us to use MI as an early stopping criterion.
    \item We extensively evaluate our method on multiple brain MRI datasets and show that it achieves high visual quality for different contrasts and views and preserves pathological details, highlighting its potential clinical usage.
\end{enumerate}

\subsubsection{Related Work.}
Single-image super-resolution (SISR) \cite{bhowmik2017training} aims at restoring a high-resolution (HR) image from a low-resolution (LR) input from a single sequence and targets applications such as low-field MR upsampling or optimization of MRI acquisition \cite{chen2018brain}.
Recent methods \cite{chen2018brain,georgescu2020convolutional} incorporate priors learned from a training set \cite{chen2018brain}, which is later combined with generative models \cite{chen2020mri}.
On the other hand, multi-image super-resolution (MISR) relies on the information from complementary views of the same sequence \cite{wu2021irem} and is especially relevant to capturing temporal redundancy in motion-corrupted low-resolution MRI \cite{gholipour2010robust,wesarg2010combining}.

Multi-contrast Super-resolution (MCSR) targets using inter-contrast priors \cite{rousseau2010non}. In conventional settings \cite{manjon2010mri}, an isotropic HR image of another contrast is used to guide the reconstruction of an anisotropic LR image.
Zeng et al. \cite{zeng2018simultaneous} use a two-stage architecture for both SISR and MCSR.
Utilizing a feature extraction network, Lyu et al. \cite{lyu2020multi} learn multi-contrast information in a joint feature space.
Later, multi-stage integration networks \cite{feng2021multi}, separatable attention \cite{feng2021exploring} and transformers \cite{li2022transformer} have been used to enhance joint feature space learning.
However, all current MCSR approaches are limited by their need for a large training dataset. Consequently, this constrains their usage to specific resolutions and further harbors the danger of hallucination of features (e.g., lesions, artifacts) present in the training set and does not generalize well to unseen data.

Originating from shape reconstruction \cite{park2019deepsdf} and multi-view scene representations \cite{mildenhall2020nerf}, Implicit Neural Representations (INR) have achieved state-of-the-art results by modeling a continuous function on a space from discrete measurements. Key reasons behind INR's success can be attributed to overcoming the low-frequency bias of Multi-Layer Perceptrons (MLP) \cite{tancik2020fourier,sitzmann2020implicit,saragadam2023wire}. 
Although MRI is a discrete measurement, the underlying anatomy is a continuous space. We find INR to be a good fit to model a continuous intensity function on the anatomical space. Once learned, it can be sampled at an arbitrary resolution to obtain the super-resolved MRI.
Following this spirit, INRs have recently been successfully employed in medical imaging applications ranging from k-space reconstruction \cite{huang2023neural} to SISR \cite{wu2021irem}.
Unlike \cite{wu2021irem,shen2022nerp}, which learn anatomical priors in single contrasts, and \cite{wu2022ASSR,amiranashvili2022learning}, which leverage INR with latent embeddings learned over a cohort, we focus on employing INR in subject-specific, multi-contrast settings.
\section{Methods}

\begin{figure}[tbh]
    \centering
    \includegraphics[width=\linewidth,trim=40 10 0 0, clip]{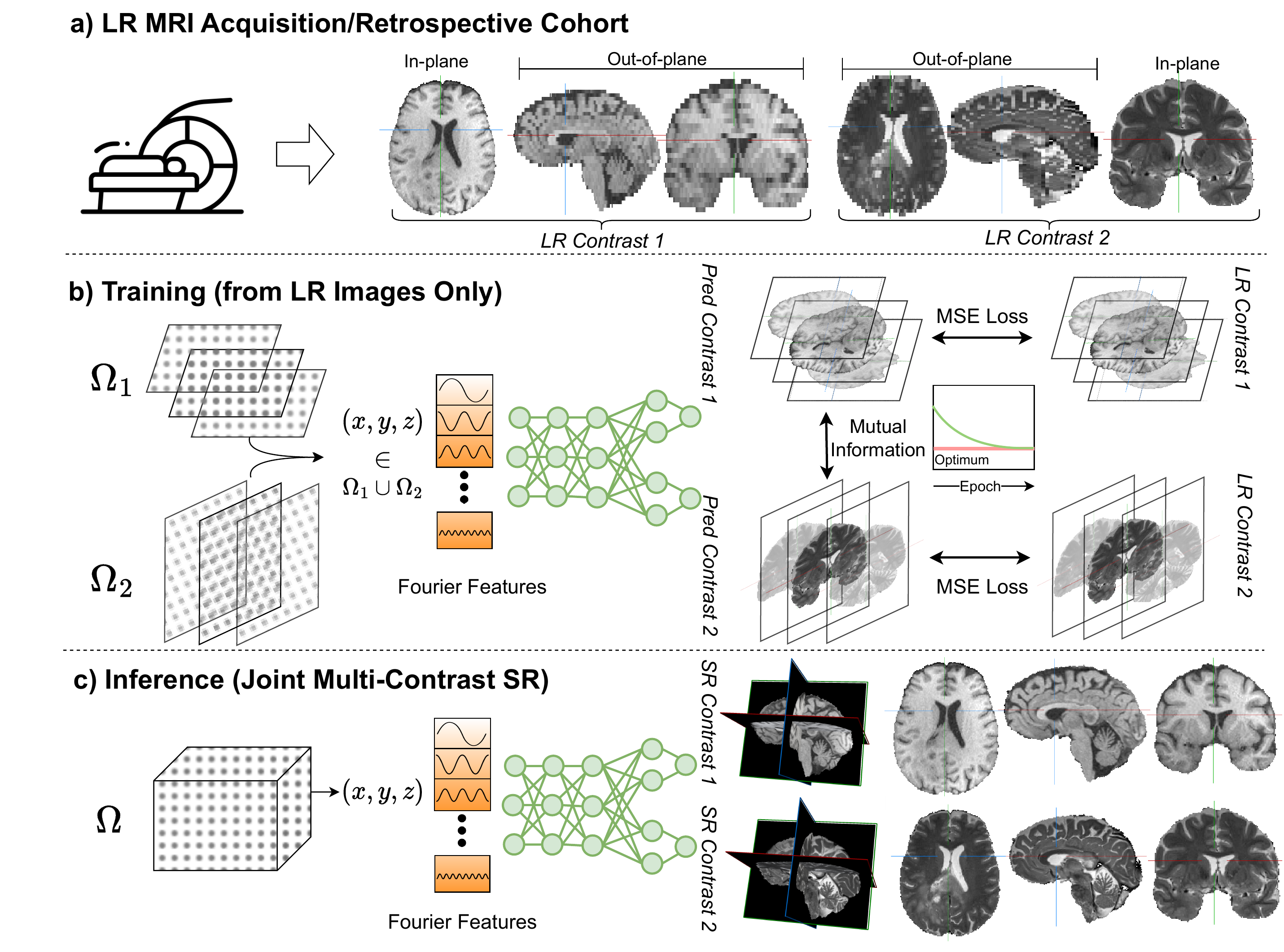}
    \caption{Overview of our proposed approach (best viewed in full screen). a) Given a realistic clinical scenario, two MRI contrasts are acquired in complementary 2D views. b) Our proposed INR models both contrast from the supervision available in the 2D scans and, by doing so, learn to transfer knowledge from in-plane measurements to out-of-plane of the other contrast. Although our model is trained on MSELoss only for the observed coordinates, it constructs a continuous function space, converging to an optimum state of mutual information between the contrasts on the global space of $\Omega$. c) Once learned, we can sample an isotropic grid and obtain the anatomically faithful and pathology-preserving super-resolution.}
    \label{fig:approach}
\end{figure}

In this section, we first formally introduce the problem of joint super-resolution of multi-contrast MRI from only one image per contrast per patient. Next, we describe strategies for embedding information from two contrasts in a shared space. Subsequently, we detail our model architecture and training configuration.\\

\noindent{\textbf{Problem Statement.}}
We denote the collection of all 3D coordinates of interest in this anatomical space as $\Omega=\{(x,y,z)\}$ with anatomical function $q:\Omega\rightarrow \mathbb{A}$.
The image intensities are a function of the underlying anatomical properties $\mathbb{A}$.
Two contrasts $C_1$ and $C_2$ can be scanned in a low-resolution subspace $\Omega_1,\Omega_2 \subset \Omega$.
Let us consider $g_1,g_2 :\mathbb{A}\rightarrow \mathbb{R}$ that map from anatomical properties to contrast intensities $C_1$ and $C_2$, respectively. We obtain 
 sparse observations $I_1=\{g_1(q(\boldsymbol{x}))=f_1(\boldsymbol{x}); \forall \boldsymbol{x} \in \Omega_1\}$ and $I_2=\{g_2(q(\boldsymbol{x}))=f_2(\boldsymbol{x}); \forall \boldsymbol{x} \in \Omega_2\}$, where $f_i$ is composition of $g_i$ and $q$.
However, one can easily obtain the global anatomical space $\Omega$ by knowing $\Omega_1$ and $\Omega_2$, e.g., by rigid registration between the two images.
In this paper, we aim to estimate $f_1,f_2:\Omega\rightarrow \mathbb{R} $ given $I_1$ and $I_2$.\\

\noindent{\textbf{Joint Multi-contrast Modelling.}} Since both component-functions $f_1$ and $f_2$ operate on a subset of the same input space, we argue that it is beneficial to model them jointly as a single function $f:\Omega\rightarrow \mathbb{R}^2$ and optimize it based on their estimation error incurred in their respective subsets. This will enable information transfer from one contrast to another, thus improving the estimation and preventing over-fitting in single contrasts, bringing consistency to the prediction.
To this end, we propose to leverage INR to model a continuous multi-contrast function $f$ from discretely sampled sparse observations $I_1$ and $I_2$.\\

\noindent{\textbf{MCSR Setup.}}
Without loss of generalization, let us consider two LR input contrasts scanned in two orthogonal planes $p_1$ and $p_2$, where $p_1, p_2\in$ \{axial, sagittal, coronal\}. We assume they are aligned by rigid registration requiring no coordinate transformation.
Their corresponding in-plane resolutions are $(s_1\times s_1)$ and $(s_2\times s_2)$ and slice thickness is $t_1$ and $t_2$, respectively. Note that $s_1<t_1$ and $s_2<t_2$ imply high in-plane and low out-of-plane resolution. In the end, we aim to sample an isotropic $(s\times s \times s)$ grid for both contrasts where $s\leq s_1,s_2$.\\

\noindent{\textbf{Implicit Neural Representations for MCSR.}}
We intend to project the information available in one contrast into another by embedding both in the shared weight space of a neural network.
However, a high degree of weight sharing could hinder contrast-specific feature learning.
Based on this reasoning, we aim to hit the sweet spot where maximum information exchange can be encouraged without impeding contrast-specific expressiveness.
We propose a split-head architecture, as shown in Fig. \ref{fig:approach}, where the initial layers jointly learn the common anatomical features, and subsequently, two heads specialize in contrast-specific information. The model takes Fourier \cite{tancik2020fourier} Features $\boldsymbol{v} = [cos(2 \pi B\boldsymbol{x}), sin(2 \pi B\boldsymbol{x})] ^{T}$ as input and predicts $[\hat{I}_1, \hat{I}_2] = f(\boldsymbol{v})$, where $\boldsymbol{x}=(x,y,z)$ and $B$ is sampled from a Gaussian distribution $\mathcal{N}(\mu, \sigma^2)$. We use mean-squared error loss, $\mathcal{L}_{{MSE}}$, for training.
\begin{equation}
    \mathcal{L}_{{MSE}}=\alpha\sum_{\boldsymbol{x}\in\Omega_1}(\hat{I}_1(\boldsymbol{x})-I_1(\boldsymbol{x}))^2+\beta\sum_{\boldsymbol{x}\in\Omega_2}(\hat{I}_2(\boldsymbol{x})-I_2(\boldsymbol{x}))^2
\end{equation}
where $\alpha$ and $\beta$ are coefficients for the reconstruction loss of two contrasts. Note that for points $\{(x,y,z)\}\in \Omega_{2} \setminus \Omega_{1}$, there is no explicit supervision coming from low resolution $C_1$. For these points, one can interpret learning $C_1$ from the loss in $C_2$, and vice versa, to be a weakly supervised task.\\

\noindent{\textbf{Implementation and Training.}} Given the rigidly registered LR images, we compute $\Omega_{1}$, $\Omega_{2} \in \Omega$ in the scanner reference space using their affine matrices. Subsequently, we normalize $\Omega$ to the interval $[-1,1]^3$ and independently normalize each contrast's intensities to $[0,1]$. We use 512-dimensional Fourier Features in the input. Our model consists of a four-layer MLP with a hidden dimension of 1024 for the shared layers and two layers with a hidden dimension of 512 for the heads. We use Adam optimizer with a learning rate of 4e-4 and a Cosine annealing rate scheduler with a batch size of 1000. For the multi-contrast INR models, we use MI as in Eq. \ref{eq:mi} for early stopping. Implemented in PyTorch, we train our model on a single A6000 GPU. Please refer to Tab. 3 in supplementary for an exhaustive hyper-parameter search. \\

\noindent{\textbf{Model Selection and Inference.}}
Since our model is trained on sparse sets of coordinates, it is prone to overfitting them and has little incentive to generalize in out-of-plane predictions for single contrast settings. A remedy to this is to hold random points as a validation set. However, this will reduce the number of training samples and hinder the reconstruction of fine details. For multi-contrast settings, one can exploit the agreement between the two predicted contrasts. Ideally, the network should reach an equilibrium between the contrasts over the training period, where both contrasts optimally benefit from each other. We empirically show that Mutual Information (MI) \cite{wells1996multi} is a good candidate to capture such an equilibrium point without the need for ground truth data in its computation. For two predicted contrasts $\hat{I}_1$ and $\hat{I}_2$, MI can be expressed as:

\begin{equation}
\mathrm{MI}(\hat{I}_1, \hat{I}_2)=\sum_{y \in \hat{I}_2} \sum_{x \in \hat{I}_1} P_{(\hat{I}_1, \hat{I}_2)}(x, y) \log \left(\frac{P_{(\hat{I}_1, \hat{I}_2)}(x, y)}{P_{\hat{I}_1}(x) P_{\hat{I}_2}(y)}\right)\label{eq:mi}
\end{equation}
Compared to image registration, we do not use MI as a loss for aligning two images; instead, we use it as a quantitative assessment metric. Given two ground truth HR images for a subject, one can compute the optimum state of MI. We observe that the MI between our model predictions converges close to such an optimum state over the training period without any explicit knowledge about it, c.f. Fig. 3 in the supplementary. This observation motivates us to detect a plateau in MI between the predicted contrasts and use it as a stopping criterion for model selection in multi-contrast INR.

\section{Experiments and Results}

\noindent{\textbf{Datasets.}}
To enable fair evaluation between our predictions and the reference HR ground truths, the in-plane SNR between the LR input scan and corresponding ground truth has to match. To synthetically create 2D LR images, it is necessary to downsample out-of-plane in the image domain anisotropically \cite{zhao2020smore} while preserving in-plane resolution. Consequently, to mimic realistic 2D clinical protocol, which often has higher in-plane details than that of 3D scans, we use spline interpolation to model partial volume and downsampling. We demonstrate our network's modeling capabilities for different contrasts (T1w, T2w, FLAIR, DIR), views (axial, coronal, sagittal), and pathologies (MS, brain tumor). We conduct experiments on two public datasets, BraTS \cite{menze2014multimodal}, and MSSEG \cite{commowick2016msseg}, and an in-house clinical MS dataset (cMS). In each dataset, we select 25 patients that fulfill the isotropic acquisition criteria for both ground truth HR scans. 
Note that we only use the ground truth HR for evaluation, not anywhere in training. We optimize separate INRs for each subject with supervision from only its two LR scans. 
If required, we employ skull-stripping \cite{isensee2019automated} and rigid registration to the MNI152 (MSSEG, cMS) or SRI24 (BraTS) templates. For details, we refer to Table 2 in the supplementary.\\

\noindent{\textbf{Metrics.}}
We evaluate our results by employing common SR \cite{dong2015image,lyu2020multi,wu2021irem} quality metrics, namely PSNR and SSIM. To showcase perceptual image quality, we additionally compute the Learned Perceptual Image Patch Similarity (LPIPS) \cite{zhang2018unreasonable} and measure the absolute error $\epsilon_{MI}$ in mutual information of two upsampled images to their ground truth counterparts as follows:
\begin{equation}
\small
    \varepsilon_{MI}^{C_i} = \frac{1}{N}\sum_{k=1}^N|\mathrm{MI}(\hat{I}_i^k, I_j^k)-\mathrm{MI}(I_i^k, I_j^k)| ;~~
    \hat{\varepsilon}_{MI} = \frac{1}{N}\sum_{k=1}^N|\mathrm{MI}(\hat{I}_1^k, \hat{I}_2^k)-\mathrm{MI}(I_1^k, I_2^k)|\nonumber
\end{equation}

\noindent{\textbf{Baselines and Ablation.}}
To the best of our knowledge, there are no prior data-driven methods that can perform MCSR on a single-subject basis. Hence, we provide single-subject baselines that operate solely on single contrast and demonstrate the benefit of information transfer from other contrasts with our proposed models. In addition, we show ablations of our proposed split head model compared to our vanilla INR. Precisely, our experiments include:

\noindent{\textbf{\emph{{Baseline 1}}:} Cubic-spline interpolation is applied on each contrast separately. \\
\noindent{\textbf{\emph{{Baseline 2}}:} LRTV \cite{shi2015lrtv} applied on each contrast separately. \\
\noindent{\textbf{\emph{{Baseline 3}}:} SMORE (v3.1.2) \cite{zhao2020smore} applied on each contrast separately. \\
\noindent{\textbf{\emph{{Baseline 4}}:} Two single-contrast INRs with one output channel each.

\noindent{\textbf{\emph{{Our vanilla INR (ablation)}}}}: Single INR with two output channels that jointly predicts the two contrast intensities.

\noindent{\textbf{\emph{{Our proposed split-head INR}}}}: Single INR with two separate heads that jointly predicts the two contrast intensities (cf. Fig. \ref{fig:approach}).\\

\begin{table}[t!]
\caption{Quantitative results for MCSR on two public and one in-house datasets. All metrics consistently show that our split-head INR performs the best for MCSR.}\label{tab1}
\begin{adjustbox}{width=\textwidth}
\begin{tabular}{l|l|r|r|r|r|r|r|r|r|r}
\toprule
\multirow{8}*{\rotatebox[origin=c]{90}{BraTS 2019}} & Contrasts &  \multicolumn{3}{c|}{T1w} & \multicolumn{3}{c|}{T2w} & \multicolumn{3}{c}{T1w \& T2w}\\
\cline{2-11}
& Methods & PSNR $\uparrow$ & SSIM $\uparrow$ & LPIPS $\downarrow$ & PSNR $\uparrow$ & SSIM $\uparrow$ & LPIPS $\downarrow$ & $\varepsilon_{MI}^{T1}$ $\downarrow$ & $\varepsilon_{MI}^{T2}$ $\downarrow$ & $\hat{\varepsilon}_{MI}$ $\downarrow$\\
\cmidrule{2-11}
& Cubic Spline & 21.201 & 0.896 & 0.098 & 26.201 & 0.932 & 0.058 & 0.096 & 0.087 & 0.145 \\
& LRTV & 21.328 & 0.919 & 0.052 & 24.206 & 0.915 & 0.053 & 0.126 & 0.127 & 0.203 \\
& SMORE & 26.266 & 0.942 & 0.030 & 28.466 & 0.942 & 0.030 & 0.157 & 0.127 & 0.225 \\

& Single Contrast INR & 26.168& 0.952& 0.036& 29.920 &0.957 & 0.028& 0.051& 0.030& 0.079\\
\cmidrule{2-11}
& Our vanilla INR & 26.196 & 0.960 & 0.032 & 29.777 & 0.962 & 0.026& 0.008& 0.015& 0.065\\
& Our split-head INR & \textbf{28.746}& \textbf{0.965}& \textbf{0.028}& \textbf{31.802} & \textbf{0.966} & \textbf{0.024}&\textbf{0.007} &\textbf{0.014} &\textbf{0.062}\\

\toprule
\multirow{8}*{\rotatebox[origin=c]{90}{MSSEG 2016}} & Contrasts &  \multicolumn{3}{c|}{T1w} & \multicolumn{3}{c|}{Flair} & \multicolumn{3}{c}{T1w \& Flair}\\
\cline{2-11}
& Methods & PSNR $\uparrow$ & SSIM $\uparrow$ & LPIPS $\downarrow$ & PSNR $\uparrow$ & SSIM $\uparrow$ & LPIPS $\downarrow$ & $\varepsilon_{MI}^{T1}$ $\downarrow$ & $\varepsilon_{MI}^{Flair}$ $\downarrow$ & $\hat{\varepsilon}_{MI}$ $\downarrow$\\
\cmidrule{2-11}
& Cubic Spline & 30.102 & 0.953 & 0.051 & 28.724 & 0.945 & 0.054 & 0.062 & 0.087 & 0.115 \\
& LRTV & 22.848 & 0.860 & 0.050 & 23.920 & 0.872 & 0.044 & 0.068 & 0.052 & 0.095 \\
& SMORE & 25.729 & 0.937 & 0.030 & 27.430 & 0.940 & 0.029 & 0.138 & 0.100 & 0.183 \\
& Single Contrast INR & 30.852  & 0.956 & 0.029 &31.156 & 0.955 & 0.030 & 0.047&0.074  & 0.095\\
\cmidrule{2-11}
& Our vanilla INR & 31.599 & 0.966 & \textbf{0.019} & 32.312 & 0.969 & 0.019 & \textbf{0.008} & 0.025 & \textbf{0.024}\\
& Our split-head INR & \textbf{31.769} & \textbf{0.967} & \textbf{0.019} & \textbf{32.514} & \textbf{0.970} & \textbf{0.018} & \textbf{0.008}& \textbf{0.023}& \textbf{0.024}\\

\toprule
\multirow{8}*{\rotatebox[origin=c]{90}{cMS}} & Contrasts &  \multicolumn{3}{c|}{DIR} & \multicolumn{3}{c|}{Flair} & \multicolumn{3}{c}{DIR \& Flair}\\
\cline{2-11}
& Methods & PSNR $\uparrow$ & SSIM $\uparrow$ & LPIPS $\downarrow$ & PSNR $\uparrow$ & SSIM $\uparrow$ & LPIPS $\downarrow$ & $\varepsilon_{MI}^{DIR}$ $\downarrow$ & $\varepsilon_{MI}^{Flair}$ $\downarrow$ & $\hat{\varepsilon}_{MI}$ $\downarrow$\\
\cmidrule{2-11}
& Cubic Spline & 28.106 & 0.929 & 0.065 & 26.545 & 0.923 & 0.079 & 0.083 & 0.096 & 0.136 \\
& LRTV & 28.725 & 0.904 & 0.033 & 22.766 & 0.835 & 0.057 & 0.269 & 0.088 & 0.312 \\
& SMORE & 28.933 & 0.926 & 0.040 & 25.336 & 0.921 & 0.039 & 0.124 & 0.079 & 0.139 \\
& Single Contrast INR & 29.941&0.937 & 0.037 & 28.655 & 0.936 & 0.041& 0.063& 0.072& 0.096\\
\cmidrule{2-11}
& Our vanilla INR & 30.816& \textbf{0.956}&0.024 &29.749 & 0.950 & 0.029 & 0.022 & \textbf{0.033} & \textbf{0.009}\\
& Our split-head INR & \textbf{31.686}& \textbf{0.956}& \textbf{0.023} & \textbf{30.246} & \textbf{0.952} & \textbf{0.028} & \textbf{0.021} & \textbf{0.033} & \textbf{0.009}\\
\bottomrule
\end{tabular}
\end{adjustbox}
\end{table}

\noindent{\textbf{Quantitative Analysis.}} Table \ref{tab1} demonstrates that our proposed framework poses a trustworthy candidate for the task of MCSR.
As observed in \cite{zhao2020smore}, LRTV struggles for anisotropic up-sampling while SMORE's overall performance is better than cubic-spline, but slightly worse to single-contrast INR. However, the benefit of single-contrast INR may be limited if not complemented by additional views as in~\cite{wu2021irem}.
For \textit{MCSR} from single-subject scans, we achieve encouraging results across all metrics for all datasets, contrasts, and views.
Since T1w and T2w both encode anatomical structures, the consistent improvement in BraTS for both sequences serves as a proof-of-concept for our approach.
As FLAIR is the go-to-sequence for MS lesions, and T1w does not encode such information, the results are in line with the expectation that there could be a relatively higher transfer of anatomical information to pathologically more relevant FLAIR than vice-versa.
Lastly, given their similar physical acquisition and lesion sensitivity, we note that DIR/FLAIR benefit to the same degree in the cMS dataset.\\

\begin{figure}[t!]
    \centering
    \includegraphics[clip, trim=7cm 4.0cm 3cm 4.0cm, scale=0.38]{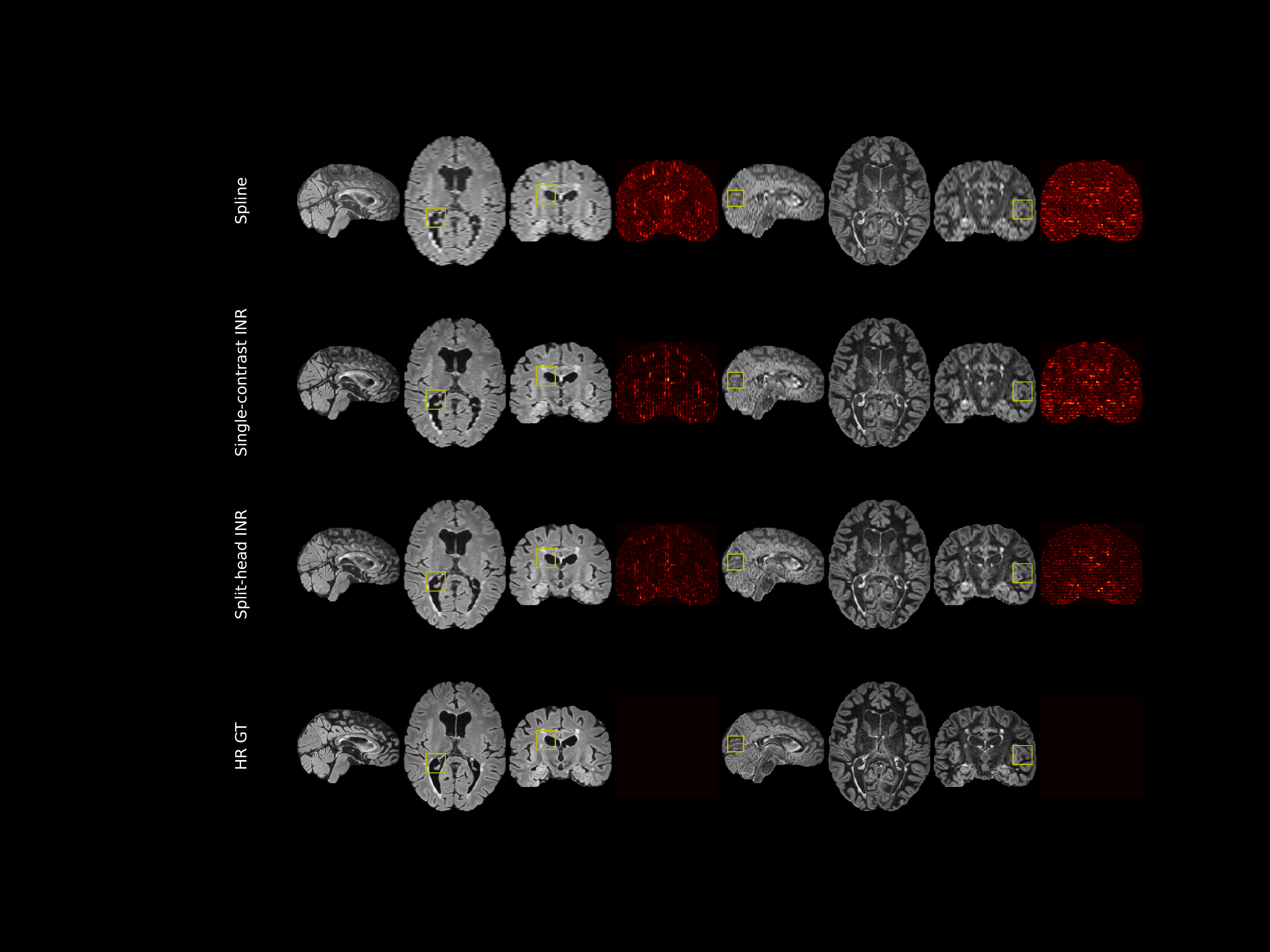}
    \caption{Qualitative results for MCSR for cMS. The predictions of the split-head INR demonstrate the transfer of anatomical and lesion knowledge from complementing views and sequences. Yellow boxes highlight details recovered by the split-head INR in the out-of-plane reconstructions, where others struggle.}
    \label{fig:visual}
\end{figure}

\noindent{\textbf{Qualitative Analysis.}}
Fig. \ref{fig:visual} shows the typical behavior of our models on cMS dataset, where one can qualitatively observe that the split-head INR preserves the lesions and anatomical structures shown in the yellow boxes, which other models fail to capture.
While our reconstruction is not identical to the GT HR, the coronal view confirms anatomically faithful reconstructions despite not receiving any in-plane supervision from any contrast during training.
We refer to Fig. 4 in the supplementary for similar observations on BraTS and MSSEG.\\

\section{Discussion and Conclusion}
Given the importance and abundance of large multi-parametric retrospective cohorts \cite{menze2014multimodal,commowick2016msseg}, our proposed approach will allow the upscaling of LR scans with the help of other sequences.
Deployment of such a model in clinical routine would likely reduce acquisition time for multi-parametric MRI protocols maintaining an acceptable level of image fidelity. Importantly, our model exhibits trustworthiness in its clinical applicability being 1) subject-specific, and 2) as its gain in information via super-resolution is validated by MI preservation and is not prone to hallucinations that often occur in a typical generative model.

In conclusion, we propose the first subject-specific deep learning solution for isotropic 3D super-resolution from anisotropic 2D scans of two different contrasts of complementary views. Our experiments provide evidence of inter-contrast information transfer with the help of INR. Given the supervision of only single subject data and trained within minutes on a single GPU, we believe our framework to be potentially suited for broad clinical applications. Future research will focus on prospectively acquired data, including other anatomies.

\subsection*{Acknowledgement} 
JM, MM and JSK are supported by Bavarian State Ministry for Science and Art (Collaborative Bilateral Research Program Bavaria – Québec: AI in medicine, grant F.4-V0134.K5.1/86/34). SS, RG and JSK are supported by European Research Council (ERC) under the European Union’s Horizon 2020 research and innovation program (101045128-iBack-epic-ERC2021-COG). MD and DR are supported by ERC (Deep4MI - 884622) and ERA-NET NEURON Cofund (MULTI-FACT - 8810003808). HBL is supported by an Nvidia GPU grant. 

%
%
%
\bibliographystyle{splncs04}
\bibliography{references}

\begin{thebibliography}{10}
\providecommand{\url}[1]{\texttt{#1}}
\providecommand{\urlprefix}{URL }
\providecommand{\doi}[1]{https://doi.org/#1}

\bibitem{amiranashvili2022learning}
Amiranashvili, T., L{\"u}dke, D., Li, H.B., Menze, B., Zachow, S.: Learning
  shape reconstruction from sparse measurements with neural implicit functions.
  In: MIDL. pp. 22--34. PMLR (2022)

\bibitem{bhowmik2017training}
Bhowmik, A., Shit, S., Seelamantula, C.S.: Training-free, single-image
  super-resolution using a dynamic convolutional network. IEEE signal
  processing letters  \textbf{25}(1),  85--89 (2017)

\bibitem{chen2020mri}
Chen, Y., Shi, F., Christodoulou, A.G., Xie, Y., Zhou, Z., Li, D.: Efficient
  and accurate {MRI} super-resolution using a generative adversarial network
  and 3d multi-level densely connected network. In: MICCAI. pp. 91--99.
  Springer (2018)

\bibitem{chen2018brain}
Chen, Y., Xie, Y., Zhou, Z., Shi, F., Christodoulou, A.G., Li, D.: Brain mri
  super resolution using 3d deep densely connected neural networks. In: ISBI.
  pp. 739--742. IEEE (2018)

\bibitem{commowick2016msseg}
Commowick, O., Cervenansky, F., Ameli, R.: {MSSEG} challenge proceedings:
  multiple sclerosis lesions segmentation challenge using a data management and
  processing infrastructure. In: MICCAI (2016)

\bibitem{dong2015image}
Dong, C., Loy, C.C., He, K., Tang, X.: Image super-resolution using deep
  convolutional networks. IEEE TPAMI  \textbf{38}(2),  295--307 (2015)

\bibitem{feng2021multi}
Feng, C.M., Fu, H., Yuan, S., Xu, Y.: Multi-contrast {MRI} super-resolution via
  a multi-stage integration network. In: MICCAI. pp. 140--149. Springer (2021)

\bibitem{feng2021exploring}
Feng, C.M., Yan, Y., Yu, K., Xu, Y., Shao, L., Fu, H.: Exploring separable
  attention for multi-contrast {MR} image super-resolution. arXiv preprint
  arXiv:2109.01664  (2021)

\bibitem{georgescu2020convolutional}
Georgescu, M.I., Ionescu, R.T., Verga, N.: Convolutional neural networks with
  intermediate loss for 3d super-resolution of ct and mri scans. IEEE Access
  \textbf{8},  49112--49124 (2020)

\bibitem{gholipour2010robust}
Gholipour, A., Estroff, J.A., Warfield, S.K.: Robust super-resolution volume
  reconstruction from slice acquisitions: application to fetal brain {MRI}.
  IEEE TMI  \textbf{29}(10),  1739--1758 (2010)

\bibitem{ha2020one}
Ha, J.Y., Baek, H.J., Ryu, K.H., Choi, B.H., Moon, J.I., Park, S.E., Kim, T.B.:
  One-minute ultrafast brain {MRI} with full basic sequences: can it be a
  promising way forward for pediatric neuroimaging? AJR  \textbf{215}(1),
  198--205 (2020)

\bibitem{huang2023neural}
Huang, W., Li, H.B., Pan, J., Cruz, G., Rueckert, D., Hammernik, K.: Neural
  implicit k-space for binning-free non-cartesian cardiac {MR} imaging. In:
  IIPMI. pp. 548--560. Springer (2023)

\bibitem{isensee2019automated}
Isensee, F., Schell, M., Pflueger, I., Brugnara, G., Bonekamp, D., Neuberger,
  U., Wick, A., Schlemmer, H.P., Heiland, S., Wick, W., et~al.: Automated brain
  extraction of multisequence {MRI} using artificial neural networks. Human
  brain mapping  \textbf{40}(17),  4952--4964 (2019)

\bibitem{li2022transformer}
Li, G., Lv, J., Tian, Y., Dou, Q., Wang, C., Xu, C., Qin, J.:
  Transformer-empowered multi-scale contextual matching and aggregation for
  multi-contrast {MRI} super-resolution. In: CVPR. pp. 20636--20645. IEEE
  (2022)

\bibitem{lyu2020multi}
Lyu, Q., Shan, H., Steber, C., Helis, C., Whitlow, C., Chan, M., Wang, G.:
  Multi-contrast super-resolution {MRI} through a progressive network. IEEE TMI
   \textbf{39}(9),  2738--2749 (2020)

\bibitem{manjon2010mri}
Manj{\'o}n, J.V., Coup{\'e}, P., Buades, A., Collins, D.L., Robles, M.: {MRI}
  superresolution using self-similarity and image priors. J. Biomed. Imaging
  \textbf{2010},  1--11 (2010)

\bibitem{menze2014multimodal}
Menze, B.H., Jakab, A., Bauer, S., Kalpathy-Cramer, J., Farahani, K., Kirby,
  J., Burren, Y., Porz, N., Slotboom, J., Wiest, R., et~al.: The multimodal
  brain tumor image segmentation benchmark (brats). IEEE TMI  \textbf{34}(10),
  1993--2024 (2014)

\bibitem{mildenhall2020nerf}
Mildenhall, B., Srinivasan, P.P., Tancik, M., Barron, J.T., Ramamoorthi, R.,
  Ng, R.: {NeRF}: Representing scenes as neural radiance fields for view
  synthesis. In: ECCV 2020. pp. 405--421. Springer (2020)

\bibitem{park2019deepsdf}
Park, J.J., Florence, P., Straub, J., Newcombe, R., Lovegrove, S.: {DeepSDF}:
  Learning continuous signed distance functions for shape representation. In:
  CVPR. pp. 165--174. IEEE (2019)

\bibitem{plenge2012super}
Plenge, E., Poot, D.H., Bernsen, M., Kotek, G., Houston, G., Wielopolski, P.,
  van~der Weerd, L., Niessen, W.J., Meijering, E.: Super-resolution methods in
  {MRI}: can they improve the trade-off between resolution, signal-to-noise
  ratio, and acquisition time? Magnetic Resonance in Medicine  \textbf{68}(6),
  1983--1993 (2012)

\bibitem{rousseau2010non}
Rousseau, F., Initiative, A.D.N., et~al.: A non-local approach for image
  super-resolution using intermodality priors. Med. Image Anal.
  \textbf{14}(4),  594--605 (2010)

\bibitem{saragadam2023wire}
Saragadam, V., LeJeune, D., Tan, J., Balakrishnan, G., Veeraraghavan, A.,
  Baraniuk, R.G.: {WIRE}: Wavelet implicit neural representations. In: CVPR.
  pp. 18507--18516. IEEE (2023)

\bibitem{shen2022nerp}
Shen, L., Pauly, J., Xing, L.: {NeRP}: implicit neural representation learning
  with prior embedding for sparsely sampled image reconstruction. IEEE TNNLS
  (2022)

\bibitem{shi2015lrtv}
Shi, F., Cheng, J., Wang, L., Yap, P.T., Shen, D.: {LRTV: MR} image
  super-resolution with low-rank and total variation regularizations. IEEE TMI
  \textbf{34}(12),  2459--2466 (2015)

\bibitem{sitzmann2020implicit}
Sitzmann, V., Martel, J., Bergman, A., Lindell, D., Wetzstein, G.: Implicit
  neural representations with periodic activation functions. In: NeurIPS. pp.
  7462--7473 (2020)

\bibitem{tancik2020fourier}
Tancik, M., Srinivasan, P., Mildenhall, B., Fridovich-Keil, S., Raghavan, N.,
  Singhal, U., Ramamoorthi, R., Barron, J., Ng, R.: Fourier features let
  networks learn high frequency functions in low dimensional domains. In:
  NeurIPS. pp. 7537--7547 (2020)

\bibitem{wells1996multi}
Wells~III, W.M., Viola, P., Atsumi, H., Nakajima, S., Kikinis, R.: Multi-modal
  volume registration by maximization of mutual information. Med. Image Anal.
  \textbf{1}(1),  35--51 (1996)

\bibitem{wesarg2010combining}
Wesarg, S., et~al.: Combining short-axis and long-axis cardiac {MR} images by
  applying a super-resolution reconstruction algorithm. In: Medical Imaging
  2010: Image Processing. vol.~7623, pp. 187--198. SPIE (2010)

\bibitem{wu2022ASSR}
Wu, Q., Li, Y., Sun, Y., Zhou, Y., Wei, H., Yu, J., Zhang, Y.: An arbitrary
  scale super-resolution approach for 3d mr images via implicit neural
  representation. IEEE JBHI  \textbf{27}(2),  1004--1015 (2023)

\bibitem{wu2021irem}
Wu, Q., Li, Y., Xu, L., Feng, R., Wei, H., Yang, Q., Yu, B., Liu, X., Yu, J.,
  Zhang, Y.: {IREM}: high-resolution magnetic resonance image reconstruction
  via implicit neural representation. In: MICCAI. pp. 65--74. Springer (2021)

\bibitem{zeng2018simultaneous}
Zeng, K., Zheng, H., Cai, C., Yang, Y., Zhang, K., Chen, Z.: Simultaneous
  single-and multi-contrast super-resolution for brain {MRI} images based on a
  convolutional neural network. Computers in biology and medicine  \textbf{99},
   133--141 (2018)

\bibitem{zhang2018unreasonable}
Zhang, R., Isola, P., Efros, A.A., Shechtman, E., Wang, O.: The unreasonable
  effectiveness of deep features as a perceptual metric. In: CVPR. pp.
  586--595. IEEE (2018)

\bibitem{zhao2020smore}
Zhao, C., Dewey, B.E., Pham, D.L., Calabresi, P.A., Reich, D.S., Prince, J.L.:
  {SMORE}: a self-supervised anti-aliasing and super-resolution algorithm for
  {MRI} using deep learning. IEEE TMI  \textbf{40}(3),  805--817 (2020)

\end{thebibliography}
\clearpage
\appendix
\title{Appendix}

\begin{table}[h!]
\centering
\caption{Summary of dataset statistics used in our experiments.}
\label{tab:dataset}
\begin{adjustbox}{width=\textwidth}
\begin{tabular}{l|c|c|c|c|c|c|c|c}
\bottomrule
\multirow{2}{*}{Dataset}  & \multicolumn{4}{c|}{Contrast-1} & \multicolumn{4}{c}{Contrast-2} \\
\cline{2-9}
 & Name & Inplane & Dim & Res. (mm) & Name & Inplane & Dim & Res. (mm) \\
 \midrule
BraTS 2019 & T1w & Axial  & (192,192,40)    & $1\times1\times4$      & T2w  & Coronal  & (192,48,160) & $1\times4\times1$   \\
MSSEG 2016 & T1w & Axial  & (160,224,40) & $1\times1\times4$   & Flair  & Sagittal  & (40,224,160) & $4\times1\times1$   \\
cMS & DIR & Axial  & (160,224,40)    & $1\times1\times4$     & Flair  & Sagittal  & (40,224,160) & $4\times1\times1$   \\
\bottomrule 
\end{tabular}
\end{adjustbox}
\end{table}

\begin{table}[h!]
\centering

\caption{To estimate the correct hyperparameters, we perform linear and grid searches on a hold-out set of subjects across all datasets. We list the sweeped hyperparameter ranges and the configurations for the final experiments.}

\label{tab:parameters}
\begin{adjustbox}{width=\textwidth}
\begin{tabular}{l|l|l}
\toprule
  Hyperparameter & Sweep Range & Final \\
 \midrule
  Fourier Features, Distribution Scale & [3.5, 5.0, step=0.1] & 4.0 \\
  Fourier Features, Scaling Factor & [0.5, 1.5, step=0.1] \& [1.0, 10, step=1.0] & 1.0 \\
  Dimension of Fourier Features & [256, 512, 1024]  & 512 \\
  Batch Size & [1000, 3000, 5000, 10000]  & 1000 \\
  Learning Rate & [1$e^{-4}$, 2$e^{-4}$, 4$e^{-4}$] & 4$e^{-4}$ \\
  Epochs & [30, 40, 50, 80, 100] & 50 \\
  Num of Layers & [4,5,6] & 5 \\
  Num of neurons & [256,512,1024,2048] & 1024 \\
\bottomrule
\end{tabular}
\end{adjustbox}
\end{table}

\begin{figure}[h!]
    \centering
    \includegraphics[width=0.48\textwidth,trim=5 0 52 17, clip]{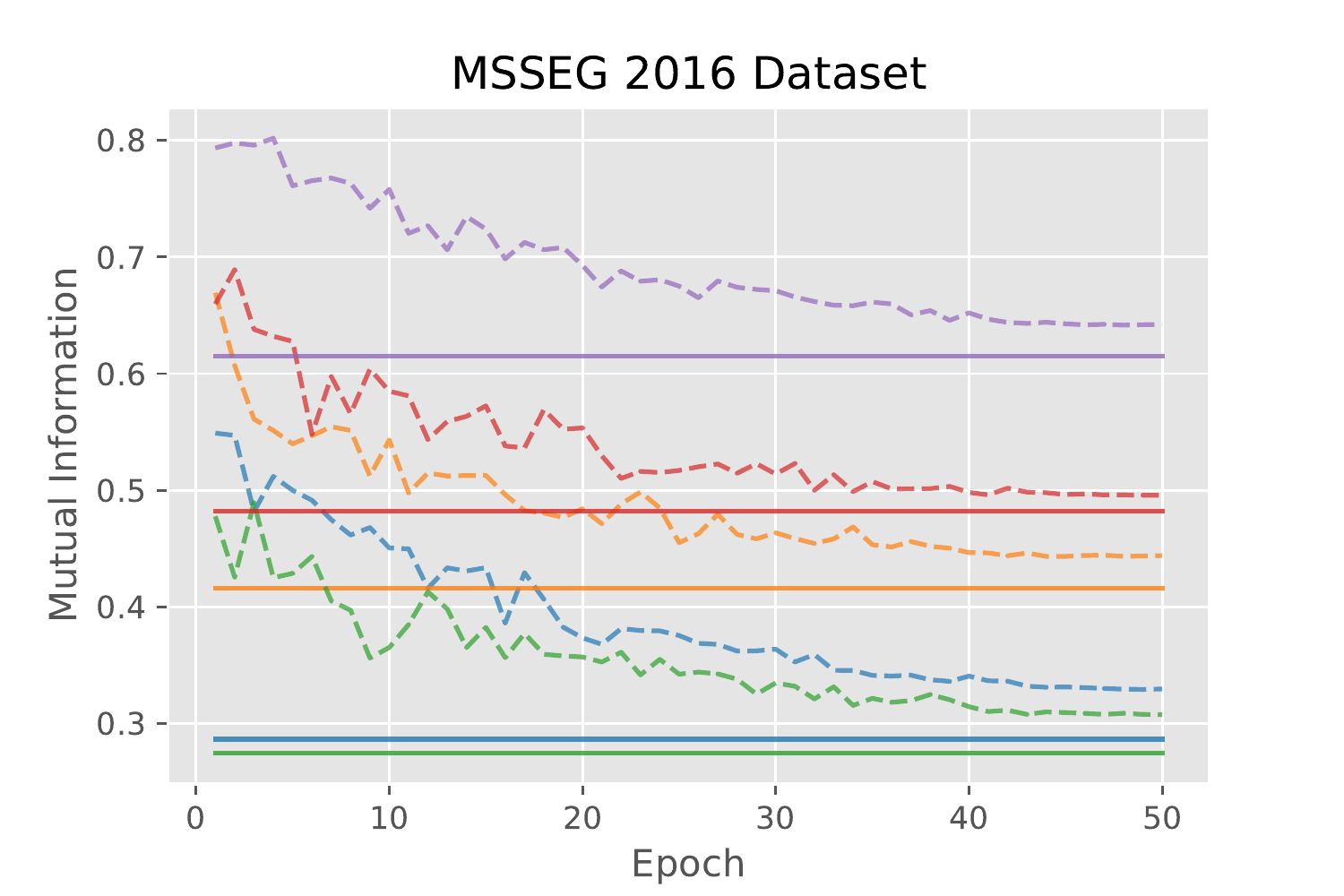}
    \includegraphics[width=0.48\textwidth,trim=5 0 52 17, clip]{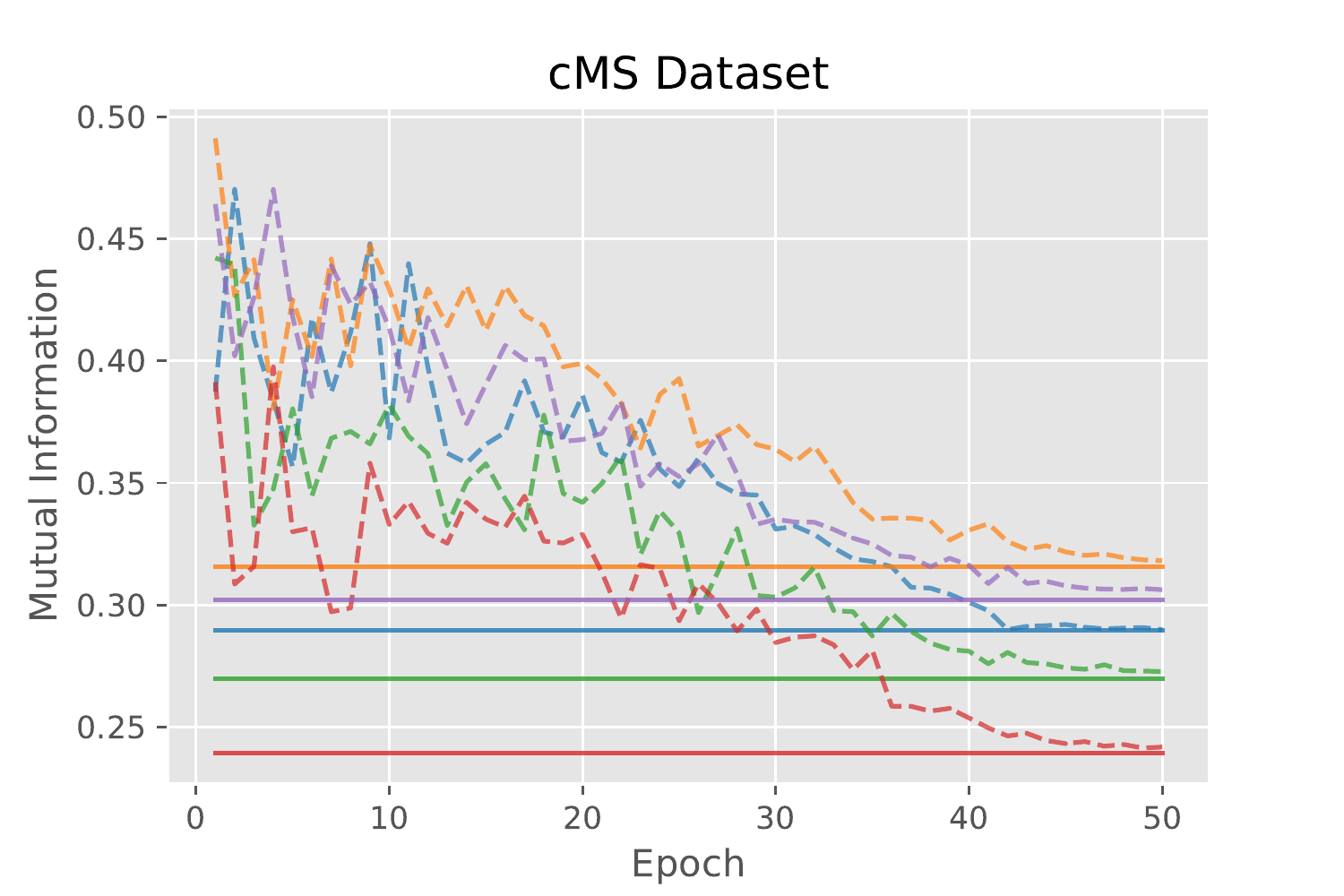}
    \caption{Convergence of predicted $MI(\hat{I}_1, \hat{I}_2)$ shown in a \textbf{dashed line} to the ground truth state $MI(I_1, I_2)$ shown in \textbf{solid line} for five randomly selected subjects (shown in a \textbf{different color}) for two datasets. Note that initially, the MI between two predicted contrasts is high because of randomly initialized shared weights, and over the training period reaches a plateau close to the true equilibrium.} 
    \label{fig:qual}
\end{figure}

\begin{figure}[tbh!]
    \centering
    \includegraphics[clip, trim=7cm 4.25cm 3cm 4.25cm, scale=0.36]{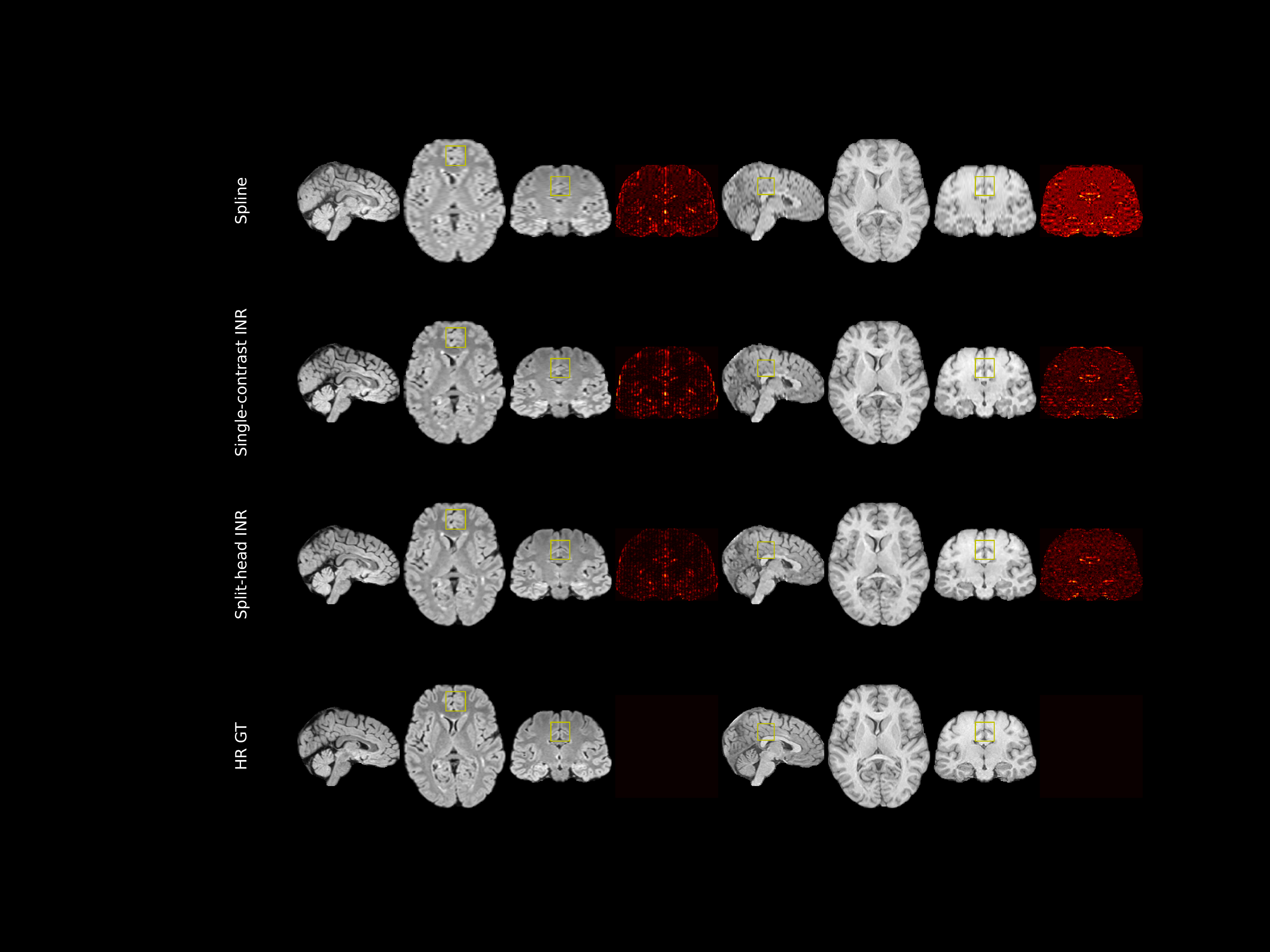}
    \includegraphics[clip, trim=7cm 4.25cm 3cm 4.25cm, scale=0.36]{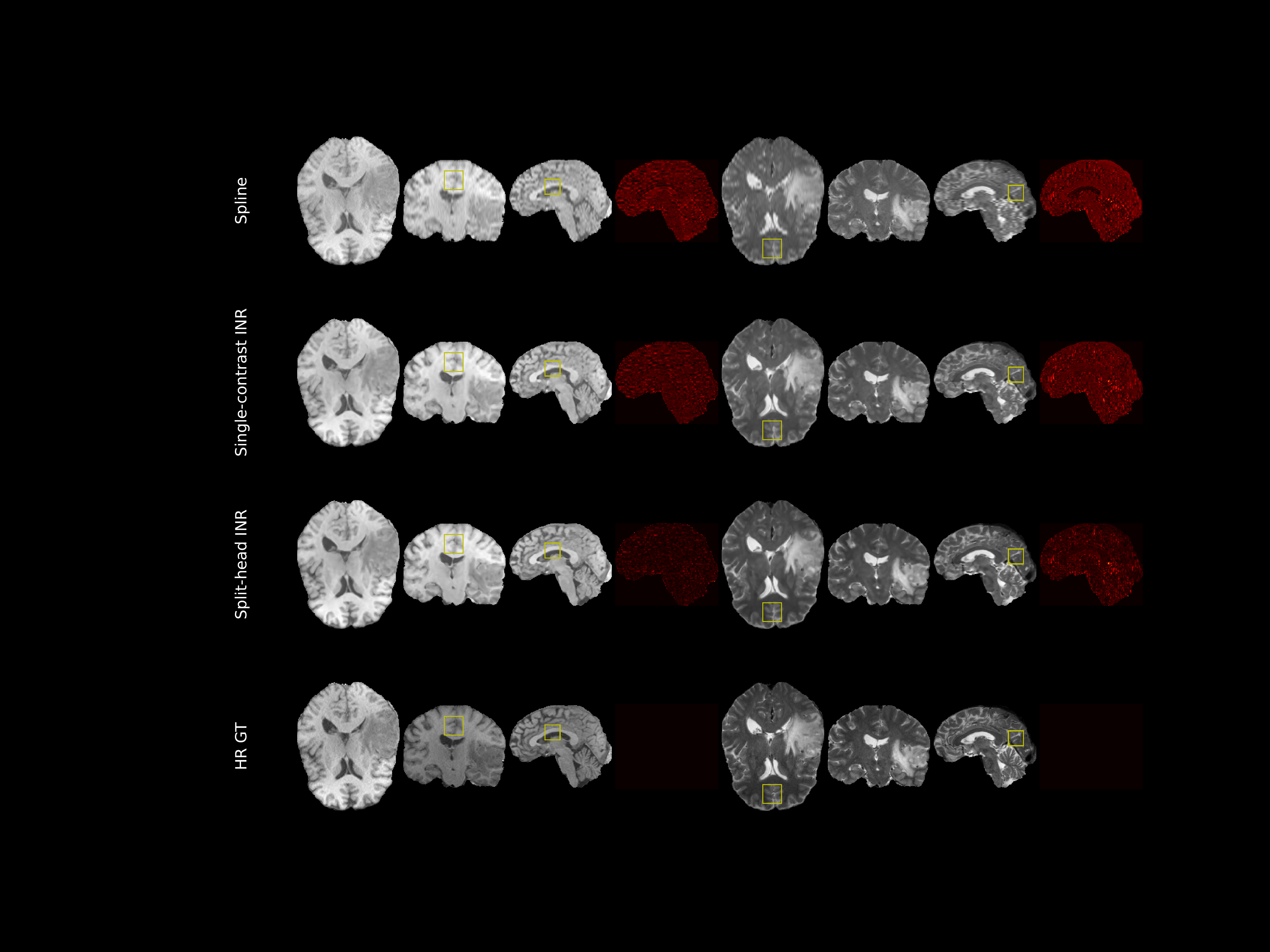}
    \caption{(Best viewed in fullscreen.) Qualitative comparisons of different models  for a typical subject of the MSSEG (upper part) and BraTS (lower part) dataset. Starting from limited out-of-plane information of the input LR scans, the split-head INR is capable of retrieving recoverable anatomical structures providing truthfulness to its prediction. 
    Exploiting the consistency and mutual anatomical information, the split-head INR can resolve ambiguities in joint reconstruction, as highlighted in yellow boxes, which is impossible if trained in a single contrast setting.
    }
    \label{fig:visual_supp}
    
\end{figure}

\end{document}